\documentclass[%
reprint,
 amsmath,amssymb,
 aps,
]{revtex4-1}

\usepackage{graphicx}
\usepackage{dcolumn}
\usepackage{bm}
\usepackage{mhchem}
\usepackage{color}
\usepackage{dcolumn}
\usepackage[T1]{fontenc}

\begin{document}

\preprint{APS/123-QED}

\title{Monoaxial Dzyaloshinskii--Moriya interaction-induced topological Hall effect in a new chiral-lattice magnet GdPt$_2$B}

\author{Yoshiki J. Sato$^{1}$}
\email{yoshiki_sato@rs.tus.ac.jp}
\author{Hikari Manako$^{1}$}
\author{Ryuji Okazaki$^{1}$}
\author{Yukio Yasui$^{2}$}
\author{Ai Nakamura$^{3}$}
\author{Dai Aoki$^{3}$}
\affiliation{%
 $^{1}$Department of Physics and Astronomy, Faculty of Science and Technology, Tokyo University of Science, Noda, Chiba 278-8510, Japan\\
 $^{2}$School of Science and Technology, Meiji University, Kawasaki, Kanagawa 214-8571, Japan\\
 $^{3}$Institute for Materials Research, Tohoku University, Oarai, Ibaraki 311-1313, Japan
}%

\date{\today}

\begin{abstract}
We investigate the topological Hall effect (THE) in the monoaxial chiral crystal GdPt$_2$B, a recently discovered compound that exhibits putative helimagnetism below 87 K. The distinct THE was observed in GdPt$_2$B in the magnetically ordered state. The scaling relations for anomalous and topological Hall conductivities differed from those of conventional models based on the scattering process. We further demonstrate the clear scaling behavior of the THE in a wide temperature range, which we attribute to the monoaxial Dzyaloshinskii--Moriya (DM) interaction under external magnetic fields perpendicular to the screw axis. The THE induced by the monoaxial DM interaction as well as the THE in a monoaxial chiral crystal of $f$-electron system are demonstrated in this study.
\end{abstract}

\maketitle

\section{Introduction}
Chirality is an important concept across many natural sciences. In physics, the absence of mirror symmetry in matter plays a key role in chirality-induced phenomena \cite{GLJAR97,BG11,KS21}. In particular, an asymmetric exchange interaction, namely the Dzyaloshinskii--Moriya (DM) interaction \cite{ID58,TM60}, arising from relativistic spin--orbit coupling (SOC), stabilizes chiral helimagnetism in magnetic materials with chiral crystal structures. The competition between an external magnetic field and magnetic interactions, including the DM interaction, induces characteristic spin textures, resulting in exotic spin-charge coupled phenomena such as the topological Hall effect (THE) \cite{AN09,NK11,NN13} and electrical magnetochiral effect \cite{TY17,RA19}.

In many studies on the magnetic and transport properties of chiral magnetic materials, B20-type cubic chiral compounds have been intensively investigated as archetypal chiral magnets \cite{SM09,AN09,XZY10,NK11,NN13,TY17}. They crystallize into the cubic space group $P2_13$ with four three-fold rotational axes. This peculiarity of multiple helical axes causes a complex interplay between the external magnetic field and the DM interaction, resulting in characteristic spin textures \cite{SM09,XZY10,NN13,TY17} and complex responses to the direction of the magnetic field \cite{WM10,NN13,TT20}. In contrast to the cubic system, hexagonal, tetragonal, and trigonal chiral crystals have one principal helical axis. They were classified as monoaxial chiral crystals.

Although a relatively large number of cubic chiral helimagnets have been reported \cite{YT21}, there are few known systems of monoaxial helimagnets with chiral crystal structures, such as  intercalated transition metal dichalcogenides (TMDs) \cite{TM83,SKK19,CZ21}, CsCuCl$_3$ \cite{YK17}, and YbNi$_3$Al$_9$ \cite{RM12}. In the monoaxial chiral crystals, the SOC causes ``monoaxial'' DM interaction along the screw axis \cite{JIK15}. The TMD magnet CrNb$_3$S$_6$ (space group $P6_322$) is an intensively investigated monoaxial helimagnet that exhibits a chiral soliton lattice (CSL) phase under a magnetic field perpendicular to the monoaxial DM interaction \cite{YT12}. The CSL exhibits strong coupling with conduction electrons and induces nontrivial magnetotransport \cite{YT13,YT16}. In addition, recent studies have shown a topological and planar Hall effect in the tilted CSL of a TMD \cite{DAM22} and transformations from a CSL to magnetic skyrmions in monoaxial TMD crystals with confined geometries \cite{LL22}. However, the question of whether the monoaxial DM interaction directly causes spin-charge coupled phenomena is an important issue. We address this issue by exploring a novel monoaxial chiral crystal system.

\begin{figure}[htbp]
\centering
\includegraphics[width=\linewidth]{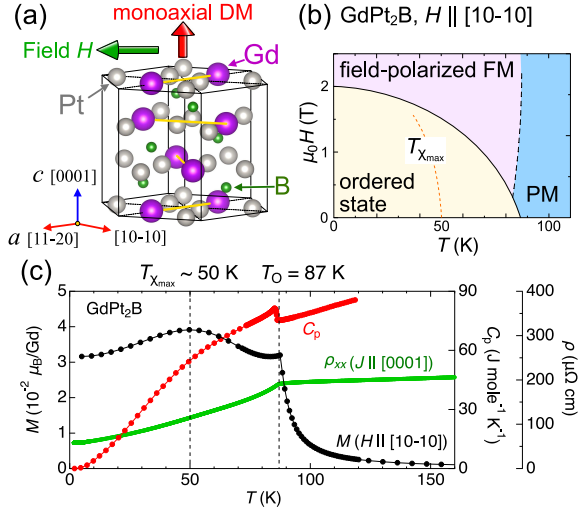}
\caption{\label{fig1} (a) Crystal structure of GdPt$_2$B. (b) Schematic magnetic phase diagram consisting of ordered state, paramagnetic (PM), and field-polarized ferromagnetic (FM) phases. (c) Temperature dependence of magnetization $M(T)$, specific heat ($C_{\rm p}$), and electrical resistivity ($\rho$). $\rho(T)$ and $C_{\rm p}(T)$ were measured in zero field, and $M(T)$ was measured in a magnetic field of $\mu_0H$ = 0.05 T. $C_{\rm p}(T)$ and $M(T)$ data were taken from Ref.\cite{YJS22}.}
\end{figure}

In this paper, we report the emergence and robust scaling behavior of the THE in a monoaxial chiral crystal GdPt$_2$B, under a magnetic field perpendicular to the screw axis. In a series of CePt$_2$B type chiral materials \cite{OS00,OLS03,RL05,MD07,RTK10,RTK15,YJS21}, the $f$-electron based magnet GdPt$_2$B is a recently discovered compound, which crystallizes in a hexagonal chiral space group $P6_422$ \cite{YJS22}. The hexagonal crystal structure and schematic magnetic ($H$-$T$) phase diagram of GdPt$_2$B are shown in Figs. \ref{fig1}(a) and \ref{fig1}(b), respectively. GdPt$_2$B is a metallic compound and exhibits putative chiral helimagnetism below $T_{\rm O}$ = 87 K. The electrical resistivity $\rho_{xx}$, specific heat $C_{\rm p}$, and magnetization $M$ exhibit a distinct phase transition at $T_{\rm O}$, as shown in Fig. \ref{fig1}(c). In addition, $M(T)$ shows a weak anomaly associated with spin reorientation at $T_{\chi_{\rm max}} \sim$ 50 K. The Curie--Weiss analysis gives a positive Weiss temperature $\theta_{\rm W} \sim 100$ K, indicating ferromagnetic exchange interaction between the Gd-ions.
A distinct THE was observed in bulk single crystals of GdPt$_2$B subjected to an external magnetic field perpendicular to the screw axis (i.e., $H$ was perpendicular to the direction of the monoaxial DM interaction). This phenomenon, arising from the competition between external magnetic fields and monoaxial DM interaction, has rarely been reported. We found the scaling behavior of the THE based on a 1-D Hamiltonian for monoaxial crystals. To the best of our knowledge, this is the first study to demonstrate the emergence of the THE induced by the monoaxial DM interaction as well as the THE in $f$-electron based monoaxial chiral crystals.

\section{Experimental details}
Single crystals of GdPt$_2$B were grown using the Czochralski method in a tetra arc furnace. The single crystals of GdPt$_2$B were oriented using a Laue camera (Photonic Science Laue X-ray CCD camera). Hall measurements were performed using bulk single crystals cut into thin plates with dimensions of 1.5 mm $\times$ 1.5 mm $\times$ 0.1 mm. The longitudinal and Hall resistances were simultaneously measured using a lock-in amplifier (SR-830) in a physical property measurement system (PPMS). Magnetization measurements were performed using a superconducting quantum interference device magnetometer (MPMS).

\section{Results and Discussion}
\begin{figure}[t]
\centering
\includegraphics[width=\linewidth]{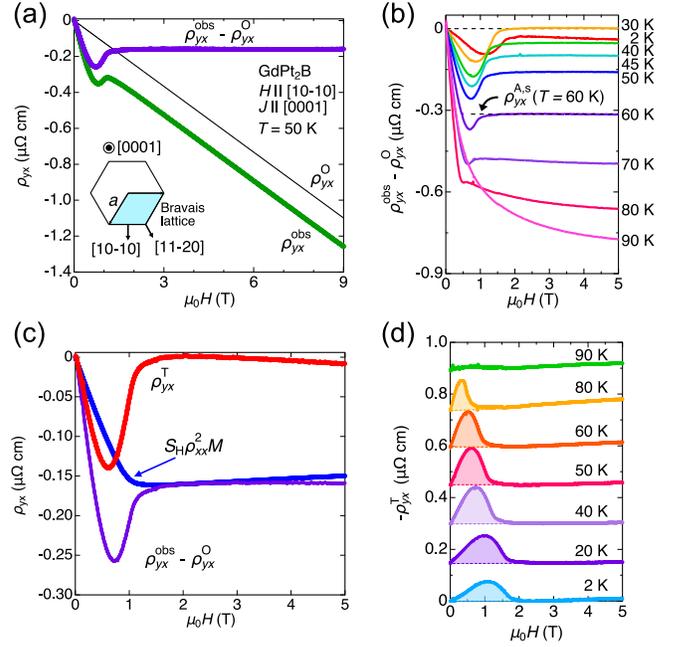}
\caption{\label{fig2} (a) Field dependence of anomalous Hall resistivity $\rho_{yx}^{\rm A}$ after subtraction of ordinary Hall resistivity $\rho_{yx}^{\rm O}$ from observed Hall resistivity $\rho_{yx}^{\rm obs}$ at 50 K. (b) $\rho_{yx}^{\rm obs} -\rho_{yx}^{\rm O}$ at several constant temperatures. (c) Topological Hall resistivity $\rho_{yx}^{\rm T}(H)$ after subtraction of anomalous Hall component derived from the magnetization ($\rho_{yx}^{\rm A}$ = $S_{\rm H}\rho_{xx}^2M$). (d) Extracted $-\rho_{yx}^{\rm T}(H)$ at several constant temperatures.}
\end{figure}

The observed Hall resistivity $\rho_{yx}^{\rm obs}$ of GdPt$_2$B for the magnetic field $H$ $||$ $[10\bar{1}0]$ and electrical current $J$ $||$ ${[0001]}$ at 50 K is shown in Fig.$~$\ref{fig2}(a). Usually, the Hall resistivity is expressed as follows:
\begin{equation}
\rho_{yx} = \rho_{yx}^{\rm O} + \rho_{yx}^{\rm A} + \rho_{yx}^{\rm T}.
\end{equation}
The first term $\rho_{yx}^{\rm O} = R_0B$ is the ordinary Hall effect (OHE) due to the Lorentz force, and $R_0$ is the ordinary Hall coefficient. The second term $\rho_{yx}^{\rm A}$ is the anomalous Hall effect (AHE) derived from the magnetization of the sample. The third term $\rho_{yx}^{\rm T}$ is the topological Hall resistivity induced by topological spin textures. In a high field region, the linear field dependence of $\rho_{yx}^{\rm obs}$ of GdPt$_2$B can be interpreted as the ordinary Hall effect $\rho_{yx}^{\rm O}$.

The field dependence of the Hall component after removing the ordinary Hall effect, namely $\rho_{yx}^{\rm obs} -\rho_{yx}^{\rm O}$, is shown in Figs. \ref{fig2}(a) and \ref{fig2}(b). A clear hump-like anomaly was observed. To extract the topological Hall component, we estimated the anomalous Hall effect corresponding to the magnetization of the sample. Generally, the anomalous Hall resistivity is described as \cite{NN10,NAP14}:
\begin{equation}
\rho_{yx}^{\rm A} = \alpha\rho_{xx}(H)M(H) + S_{\rm H}\rho^2_{xx}(H)M(H),
\end{equation} 
where $\alpha$ and $S_{\rm H}$ are material specific constants. The first and second terms correspond to skew scattering and intrinsic (Berry phase) contributions, respectively. As shown in Fig.$~$\ref{fig2}(c), the field dependence of $\rho_{yx}^{\rm A}$ can be reasonably explained by the intrinsic contribution $S_{\rm H}\rho^2_{xx}(H)M(H)$ in the field-polarized FM state. Here, the $M(H)$ data were corrected using the demagnetizing factor $D_z$ $\sim$ 0.8 for precise analysis \cite{AA98} (see Supplemental Material (SM) for details \cite{SM}). The fact that $\rho_{yx}^{\rm A}$ can be well explained by assuming an intrinsic contribution above $T_{\rm O}$ indicates the validity of the current analysis (see SM \cite{SM}).

The extracted topological Hall resistivity, namely $\rho_{yx}^{\rm T}(H) = \rho_{yx}^{\rm obs}(H) - \rho_{yx}^{\rm O}(H) - \rho_{yx}^{\rm A}(H) $, is shown in Figs.$~$\ref{fig2}(c) and \ref{fig2}(d). The hump-like field dependence of $\rho_{yx}^{\rm T}$ resembles the Hall effect induced by topological spin textures such as skyrmion lattices in B20 chiral helimagnets \cite{AN09,NK11,MK18}. The topological Hall resistivity of GdPt$_2$B is negative and reaches a maximum value of $|\rho_{yx}^{\rm T}| = 0.14$ ${\rm \mu \Omega\,cm}$ at $T$ = 45 K. The order of magnitude of $\rho_{yx}^{\rm T}$ is comparable to the reported values of topological Hall resistivity of MnGe ($\sim$ 0.16 ${\rm \mu \Omega\,cm}$ \cite{NK11}) and EuPtSi ($\sim$ 0.12 ${\rm \mu \Omega\,cm}$ \cite{MK18}). As shown in Fig.$~$\ref{fig2}(d), the hump-like anomaly in $\rho_{yx}^{\rm T}$ vanishes above the transition temperature $T_{\rm O}$, suggesting that the magnetically ordered state and the spin chirality are closely related to the emergence of the THE in GdPt$_2$B.

\begin{figure*}[htbp]
\centering
\includegraphics[width=\linewidth]{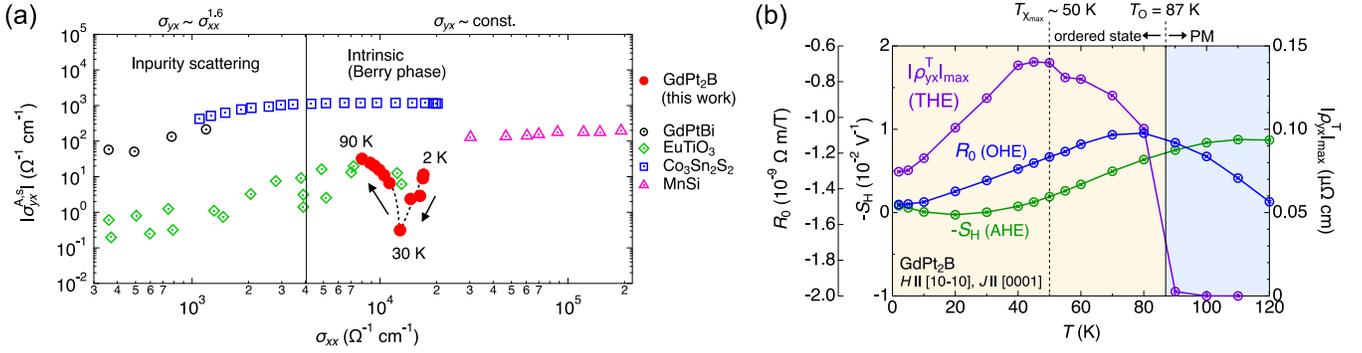}
\caption{\label{fig3}(a) Scaling relations between the Hall conductivity $|\sigma_{yx}^{\rm A,s}|$ and the longitudinal conductivity $\sigma_{xx}$ for GdPt$_2$B and several compounds \cite{TS16,KST18,EL18}. (b) Temperature dependence of coefficients $R_0$, $-S_{\rm H}$, and maximum value of $|\rho_{yx}^{\rm T}|$.}
\end{figure*}

Here, we discuss the scaling relation for AHE between $\sigma_{yx}^{\rm A}$ and $\sigma_{xx}$ arising from the scattering mechanism. Figure \ref{fig3}(a) shows a log-log plot of $|\sigma_{yx}^{\rm A,s}|$ versus $\sigma_{xx}$. $\sigma_{yx}^{\rm A,s}$ is defined as ($-\rho_{yx}^{\rm A,s}$)/[$(\rho_{yx}(H_{\rm s})^2 + \rho_{xx}(H_{\rm s})^2$], where $\rho_{yx}^{\rm A,s}$ is the saturation value of $\rho_{yx}^{\rm obs} -\rho_{yx}^{\rm O}$ in a field of $H_{\rm s}$ [see Fig. \ref{fig2}(b)]. $\sigma_{xx}$ is calculated as [$\rho_{xx}(H_{\rm s})$]/[$\rho_{yx}(H_{\rm s})^2 + \rho_{xx}(H_{\rm s})^2$]. In a "bad metal" regime with a small longitudinal conductivity, the scaling law $\sigma_{yx}$ $\propto$ $\sigma_{xx}^{1.6}$ has been demonstrated in several materials \cite{WLL04,TM07,SO08,KST18}. In contrast, $\sigma_{yx}$ becomes nearly constant in the intermediate regime ($\sigma_{xx}$ $\sim$ 10$^4$ to 10$^6$ ${\rm \Omega^{-1}cm^{-1}}$ \cite{TM07,SO08}). The data for GdPt$_2$B and several other compounds are plotted in Fig. \ref{fig3}(a). The scaling plot suggests that GdPt$_2$B is in the intermediate regime, but the scaling behavior of the AHE in GdPt$_2$B exhibits a peculiar temperature dependence.

At the lowest temperature (2 K), $|\sigma_{yx}^{\rm A,s}|$ is approximately 12 ${\rm \Omega^{-1}cm^{-1}}$ with a saturation magnetic moment $M_{\rm sat}$ of 7.2 ${\rm \mu_{\rm B}}$/Gd. As the temperature increases, $|\sigma_{yx}^{\rm A,s}|$ shows a minimum value of $|\sigma_{yx}^{\rm A,s}|$ = 0.2 ${\rm \Omega^{-1}cm^{-1}}$ ($M_{\rm sat}$ = 6.5 ${\rm \mu_{\rm B}}$/Gd) at 30 K and reaches a nearly constant value of $|\sigma_{yx}^{\rm A,s}| \sim$25 ${\rm \Omega^{-1}cm^{-1}}$ ($M_{\rm sat}$ = 4.3 ${\rm \mu_{\rm B}}$/Gd) above 60 K. $M_{\rm sat}$ and magnetoresistance vary systematically in the high field region. Therefore, the temperature dependence of the AHE in GdPt$_2$B cannot be explained by conventional AHE scenarios. The breakdown of the scaling relation of AHE is an interesting phenomenon. The mechanisms such as electron-phonon scattering\cite{CX19} and Kondo coherence in an $f$-electron system\cite{HS23} have recently been discussed as the origin of the breakdown of the scaling relation. The mechanism underlying the peculiar scaling behavior of AHE in GdPt$_2$B remains unclear.

Figure \ref{fig3}(b) shows the temperature dependence of the coefficients $R_0$ and $-S_{\rm H}$, together with the maximum value of topological Hall resistivity $|\rho_{yx}^{\rm T}|_{\rm max}$. Whereas, the THE of GdPt$_2$B is rapidly suppressed in the PM and field-polarized FM states, and the OHE and AHE exhibit a smooth temperature dependence near $T_{\rm O}$. $R_0$ was negative at all measured temperatures, indicating that the electron is a dominant carrier. The charge carrier density $n$ is estimated to be $n$ = 5.74$\times$10$^{27}$ m$^{-3}$ and $n$ = 4.20$\times$10$^{27}$ m$^{-3}$ at 80 and 2 K, respectively. $R_0$ takes its maximum value near $T_{\rm O}$. The coefficient $-S_{\rm H}$ exhibits a weak temperature dependence near $T_{\rm O}$ and reaches its maximum value at $T$ $=$ 110 K. $|\rho_{yx}^{\rm T}|_{\rm max}$ peaks near $T_{{\chi}_{\rm max}}$ $\sim$ 50 K.

\begin{figure*}[htbp]
\centering
\includegraphics[width=\linewidth]{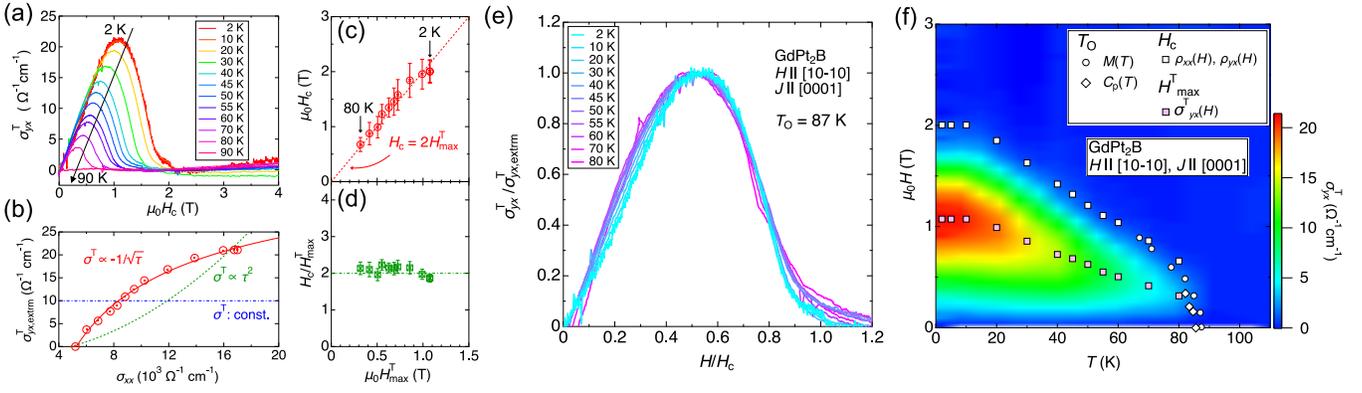}
\caption{\label{fig4}(a) Topological Hall conductivity $\sigma_{yx}^{\rm T}$ as a function of magnetic field. (b) Scaling plot of extreme value of $\sigma_{yx}^T(H,T)$ as a function of $\sigma_{xx}(H,T)$. (c) The critical field $H_{\rm c}$ as a function of the magnetic field $H^{\rm T}_{\rm max}$, which corresponds to $\sigma_{yx,{\rm extrm}}^{\rm T}$. (d) The ratio of $H_{\rm c}$ to $H^{\rm T}_{\rm max}$. (e) Scaled topological Hall conductivity $\sigma_{yx}^{\rm T}$/$\sigma_{yx,{\rm extrm}}^{\rm T}$  as a function of the scaled external magnetic field $H$/$H_{\rm c}$. (f) Contour plot of $\sigma_{yx}^{\rm T}$ in the magnetic phase diagram of GdPt$_2$B for $H$ || $[10\bar{1}0]$.}
\end{figure*}

In order to elucidate the origin of THE in GdPt$_2$B, we examined the field dependence of topological Hall conductivity $\sigma_{yx}^{\rm T}$ for $H$ || $[10\bar{1}0]$ at several constant temperatures, as shown in Fig.$~$\ref{fig4}(a). $\sigma_{yx}^{\rm T}$ is calculated as [$-\rho_{yx}^{\rm T}(H)$]/[$\rho_{yx}(H)^2 + \rho_{xx}(H)^2$]. The maximum value of $\sigma_{yx}^{\rm T}$ reaches 21 ${\rm \Omega^{-1}cm^{-1}}$ at 2 K and decreases with increasing temperature. 

The THE exhibits different scattering time $\tau$ dependence depending on its mechanism, in conventional scenarios\cite{MO04,NK11,YS12,NV22}. In the case of the intrinsic mechanism due to the momentum-space Berry curvature, the anomalous Hall component is independent of $\tau$ \cite{TM07,SO08}. In contrast, $\sigma^{\rm T}$ is expected to show $\tau^2$ dependence in the case of the THE induced by topological spin textures such as magnetic skyrmions because a fictitious magnetic flux in real space is proportional to the number of topological spin textures in a given area \cite{MO04,PB04,NV22}. We plot extreme values of $\sigma_{\rm yx}^{\rm T}$ as a function of $\sigma_{xx}$ in Fig.$~$\ref{fig4}(b), where $\sigma_{xx}$ is proportional to $\tau$ under the assumption that $n$ and the effective mass of the carrier are constant. The $\tau$ dependence of the topological Hall conductivity of GdPt$_2$B differs from those of conventional models based on momentum-space and real-space Berry curvatures. From the power-law fitting, we found that the extreme value of $\sigma_{\rm yx}^{\rm T}$ is proportional to $1/\sqrt{\tau}$ in GdPt$_2$B. A previous theoretical study explored the $\tau$ dependence of the length of the topological spin texture $L_{\rm s}^\ast$, which correspond to the length where $\sigma_{yx}^{\rm T}$ reaches its extreme value \cite{AM21}. For short-pitch skyrmion lattices, it was verified that $L_{\rm s}^\ast$ is proportional to $\sqrt{\tau}$ instead of the conventional $\tau$ dependence because of the band separation in the magnetic Brillouin zone \cite{AM21}. The nontrivial $1/\sqrt{\tau}$ dependence of the THE in GdPt$_2$B may be attributed to the specific period and configuration of the spin texture.

The most salient feature of the THE in GdPt$_2$B is that its mechanism is closely related to the monoaxial DM interaction. To discuss the connection between the THE and monoaxial DM interaction, we first define $H^{\rm T}_{\rm max}$ as the magnetic field corresponding to the maximum value of $\sigma_{yx}^{\rm T}$. Figure \ref{fig4}(c) shows the critical field of the ordered state $H_{\rm c}$ as a function of $H^{\rm T}_{\rm max}$, exhibiting a linear relationship. Thus, the ratio of $H_{\rm c}$ to $H^{\rm T}_{\rm max}$ remained constant, as shown in Fig. \ref{fig4}(d). Remarkably, the relationship between $H_{\rm c}$ and $H^{\rm T}_{\rm max}$ holds for the entire temperature range below $T_{\rm O}$.

To discuss the origin of the constant ratio $H_{\rm c}$/$H^{\rm T}_{\rm max}$, we consider a simple 1-D Hamiltonian \cite{JIK15} for monoaxial crystal systems as follows:
\begin{equation}
\mathcal{H} = -J\sum_{i} \textbf{\textit{S}}_i\cdot \textbf{\textit{S}}_{i+1} - \textbf{\textit{D}}\cdot \sum_i \textbf{\textit{S}}_i \times \textbf{\textit{S}}_{i+1} + {\textbf{\textit{H}}}\cdot\sum_i \textbf{\textit{S}}_i,
\end{equation}
where $\textbf{\textit{S}}_i$ ($|\textbf{\textit{S}}_i| = S$) is the local moment at site $i$, $J$ is the ferromagnetic exchange interaction between two adjacent spins, and $\textbf{\textit{D}}$ ($|\textbf{\textit{D}}| = D$) is the monoaxial DM interaction along the screw axis. The third term represents the effect of the external magnetic field perpendicular to the screw axis. Based on the above 1-D Hamiltonian, Kishine and Ovchinnikov derived $H_{\rm c}$ as follows \cite{JIK15}:
\begin{equation}
\label{eq4}
H_{\rm c} = \left(\frac{\pi}{4}\right)^2S\frac{D^2}{J}.
\end{equation}
Equation (\ref{eq4}) indicates that the $H_{\rm c}$ is a measure of the ratio of monoaxial DM interaction $D$ and the ferromagnetic exchange interaction $J$. Therefore, the constant ratio of $H_{\rm c}$/$H^{\rm T}_{\rm max}$ suggests that the THE of GdPt$_2$B attains its maximum values at a specific ratio of $D^2/J$ and the external magnetic field perpendicular to the screw axis. The observed THE could be scaled by the critical field $H_{\rm c}$. Figure$~$\ref{fig4}(e) shows the normalized topological Hall conductivity $\sigma_{yx}^{\rm T}$/$\sigma_{yx,{\rm extrm}}^{\rm T}$ as a function of the scaled external magnetic field $H$/$H_{\rm c}$. The observed robust scaling behavior indicates that the monoaxial DM interaction plays a significant role in the emergence of THE in the monoaxial chiral crystal GdPt$_2$B.

To summarize the results of the THE in GdPt$_2$B, we show a contour plot of the topological Hall conductivity in the $H$-$T$ phase diagram, as shown in Fig.$~$\ref{fig4}(f). An important feature is that the THE of GdPt$_2$B is stable at the lowest temperature (2 K), without the influence of thermal agitations as observed in the DM-mediated skyrmions in chiral magnets. A distinct THE is observed only within the ordered state, and the extreme value of $\sigma_{\rm yx}^{\rm T}$ depends on the scattering time. The results indicate that the spin texture in the ordered state is related to the emergence of the THE. In addition, $H_{\rm c}$ and $H^{\rm T}_{\rm max}$ show a constant ratio below $T_{\rm O}$, suggesting the importance of the monoaxial DM interaction in the THE. It is noteworthy that the scalar spin chirality [$\textbf{\textit{S}}_i \cdot (\textbf{\textit{S}}_j \times \textbf{\textit{S}}_k)$] vanishes in coplanar spin configurations such as the simple helimagnetism and CSL. Recent studies have discussed novel mechanisms such as the noncolinear Hall effect in coplanar spin configurations \cite{JB21}. Magnetic structure analysis is a desirable next step in determining the microscopic origin of the monoaxial DM interaction-induced THE in GdPt$_2$B.

Finally, we compare the THE in GdPt$_2$B with those in other material systems. In Gd-based centrosymmetric magnets such as Gd$_2$PdSi$_3$ \cite{SRS99,TK19} and Gd$_3$Ru$_4$Al$_{12}$ \cite{MH19}, a large THE induced by the magnetic frustration in a hexagonal crystal has been reported. The Ruderman--Kittel--Kasuya--Yosida (RKKY) interaction produces modulated spin textures including the skyrmion lattice with three spiral spin modulations, that is, the triple-$\textbf{\textit{q}}$ state. Owing to the complex magnetic interactions, the THE in Gd-based frustrated magnets is observed only in the intermediate magnetic phase, which is different from the present case for GdPt$_2$B. The dominant exchange interaction is considered to be of the RKKY type in GdPt$_2$B. The comparison of the THE in Gd-based monoaxial chiral crystal with those in centrosymmetric Gd-based magnets is important for a better understanding of the role of the DM interaction and mechanisms of THE. The present result is the first observation of the THE in a monoaxial chiral crystal of $f$-electron system, to the best of our knowledge. Another interesting example is large THE \cite{DAM22} and AHE \cite{NJG18} in the intercalated TMDs under the external magnetic fields along the helical axis. In the monoaxial chiral helimagnet CrNb$_3$S$_6$, a large THE induced by a tilted CSL state has been reported for $H$ $||$ $c$ (helical axis). In general, the magnetic field along the helical axis leads to a simple conical structure; however, the monoaxial DM interaction may be effective for $H$ $||$ helical axis. The anisotropy of the THE in monoaxial chiral crystals needs to be clarified in the future.
 
\section{Conclusions}
In conclusion, the large and distinct THE was observed in the monoaxial chiral crystal GdPt$_2$B. The characteristic $\tau$ dependence of the extreme value of $\sigma_{\rm yx}^{\rm T}$ and the constant ratio of $H_{\rm c}$/$H^{\rm T}_{\rm max}$ lead to the clear scaling behavior of the topological Hall conductivity, indicating that the THE in GdPt$_2$B was induced by the monoaxial DM interaction under external magnetic fields. The THE induced by the monoaxial DM interaction and THE in a monoaxial chiral crystal of $f$-electron system were demonstrated in this study. Our findings extend the range of material system in which the THE can be observed, and demonstrate the universal nature of the spin-charge coupled phenomena in topological materials.

\begin{acknowledgments}
We would like to thank S. Ohara, S. Nakamura, and Y. Yamane for discussions. This work was supported by JSPS KAKENHI (JP22K20360). We acknowledge all support from the International Research Center for Nuclear Materials Science at Oarai (IMR, Tohoku University).
\end{acknowledgments}

\bibliography{GdPt2B_Hall}

\end{document}